# Characterizing the Effect of Electrode Shift & Sensor Reapplication on Common sEMG Features in Lower Limb Muscles

Fraser Douglas, Mona Pei, Calvin Kuo

***Abstract*—This study investigates the impact of electrode shift and sensor reapplication on common surface electromyography (sEMG) features in lower limb muscles, factors which have, thus far, precluded clinicians from being able to attribute inter-session changes in sEMG signal properties to physiological changes in patients under the context of stroke recovery monitoring. To explore these inter-session errors, we recruited 12 healthy participants to perform a selection of isometric and dynamic exercises seen within stroke assessment sessions while instrumented with high-density sEMG (HDsEMG) arrays on the gastrocnemius medialis, tibialis anterior, semitendinosus, and tensor fascia latae. Between exercise sets, the electrode arrays were intentionally shifted and reapplied to quantify errors in signal features, using 3D scanning equipment to extract the ground truth shift performed. Results revealed that while frequency-domain features (mean, median, and peak frequency) demonstrated high resilience to the inter-session changes, the time-domain features (integrated EMG and max envelope amplitude) showed a greater, yet predictable, variability. In all, these findings suggest that should we be able to quantify placement shift, this can support direct inter-session feature comparisons, improving the reliability of sEMG-based stroke recovery assessments and offering insights for improving remote stroke rehabilitation technologies.**

## I. INTRODUCTION

ALTHOUGH surface electromyography (sEMG) has existed for over a century, its commercial and clinical viability only emerged in the mid-1980s [1],[2]. Since then, its use continues to grow exponentially, particularly in the fields of neurorehabilitation, assistive and prosthetic devices, and sports performance [3]. In stroke rehabilitation, sEMG has been used to quantify muscle activity and coordination to assess and classify impairment [4]-[6]. Traditionally, when assessing impairment, these regularly-attended sessions would occur in-person, although COVID-19 has catalyzed a transition to remote assessments held via videoconferencing platforms [7]. While this has been beneficial to many given their increased accessibility, particularly for those living in remote areas, they: 1) preclude clinicians from palpating the patients' muscles to assess their tone and activity, and 2) depend heavily on webcam quality and internet stability to accurately infer exercise form, which has the potential to increase subjectivity when evaluating the patient's physical condition [8].

It is for this reason that the push to integrate more objective methods of remote recovery assessment into the stroke rehabilitation process has never been greater [9], [10]. In response, we have seen a concerted effort to further incorporate wearable sensing within the field [11], where wearable sEMG measurements can serve to supplement current videoconference-based assessments with quantitative metrics supporting the efficacy of their rehabilitative interventions. Thus far in the wearable sensing field, we have seen inertial measurement units (IMUs) applied in continuous activity monitoring, linking increased physical activity to improved stroke recovery outcomes [12]-[16], quantifying balance impairments [17]-[19], and measuring gait parameters - something critical to the goal of remote stroke recovery assessment where gait recovery is often a priority [20]-[23].

Despite their merits, on their own IMUs measure only motion. This is a key limitation given that muscle control is an under constrained problem wherein multiple different muscles can activate to produce the same movement, making it difficult to assess using IMUs alone whether compensatory strategies are being employed to achieve the requested movements during stroke assessment sessions. In response, the direct measurement of muscle activity through sEMG can prove invaluable, ensuring that it is indeed the targeted muscles that are activated to achieve the desired movement.

Thus far in stroke rehabilitation, sEMG has been used primarily to drive technological interventions such as neuromuscular stimulation or exoskeleton control [24]-[27]. While these applications are not necessarily well suited to remote use, other studies have demonstrated that sEMG's ability to extract metrics relating to muscle activity has the potential to help clinicians track recovery and inform rehabilitation recommendations [28]-[32]

Now, while these measures are certainly useful within single sessions, sEMG's signal properties are highly sensitive to

This work was supported in part by a Canada Foundation for Innovation John R. Evans Leaders Fund Grant and a Canadian Natural Sciences and Engineering Research Council Discovery Grant awarded to Calvin Kuo; as well as an AGE-WELL Graduate Student Award to Fraser Douglas.

Fraser Douglas, Mona Pei, and Calvin Kuo are with the Human Motion Biomechanics Lab (HuMBL) as part of the School of Biomedical Engineering at the University of British Columbia, 2222 Health Sciences Mall, Vancouver, BC V6T 1Z3 (e-mail: fraser.douglas@ubc.ca, peimeng@student.ubc.ca, calvin.kuo@ubc.ca)

changes in sensor placement and the skin-electrode interface (affected by how the skin and sensing instruments are prepared) [33]-[35]. As a result, sEMG has seen limited clinical adoption due to the expertise required for its setup, use, and signal interpretation thereafter [36], [37]. To try and mitigate these effects, the use of the SENIAM placement and skin preparation guidelines are widely recommended [38]. However, despite these recommendations, palpation and surface anatomical landmarks are often insufficient to ensure consistent placement across multiple recording sessions, particularly if the patient is responsible for the sensor's placement in the use-case of remote assessments, thus precluding clinicians from associating changes in signal properties to physiological changes in the patient.

This idea of placement variation or electrode shift is an issue discussed across sEMG literature, and sEMG control research has demonstrated the ability of machine learning algorithms to detect if a shift has occurred and recalibrate to achieve the same level of performance as pre-shift [39], [40], these methods are not sufficient to quantify how the features used to inform patient recovery in a clinical context might have changed session to session. In response, there is a need to decouple the effects of shift from inter-session effects given the variations that arise in the skin-electrode interface. Thus, we pose the research questions of: 1) can the difference in sEMG feature values be expressed as a function of distance, 2) what is the error induced by removing and reapplying the sensor array, and 3) how do they combine?

## II. METHODS

To decouple the sEMG feature changes due to inter-session variabilities, we recruited 12 healthy participants (4 male, 8 female, mean age 28.75, standard deviation 5.9) with informed consent under the University of British Columbia Research Ethics Board H22-02423

### A. Instrumentation

For each participant, data were collected from two shank muscles - gastrocnemius medialis (GM), and tibialis anterior (TA) as well as two thigh muscles - semitendinosus (ST) and the tensor fascia latae (TFL). These were chosen due to the fundamental role they play in stroke gait compensations and recovery when compared to healthy subjects [41]-[44]. Participants were given the option to opt out of ST and/or TFL instrumentation if they were uncomfortable with having their thigh instrumented.

For each muscle on the dominant leg we performed standard skin preparation by shaving excess hair from the recording sites, abrading the areas using Nurep Skin Prep Gel to remove any dead skin cells, before wiping the areas clean and then marking the muscles' bellies with reference to SENIAM recommendations [38]. A ground electrode was attached to the electrically neutral tissue at both the ankle for the shank muscles, and the knee for thigh muscles. Both these reference sites underwent the same skin preparation.

After being prepared with their accompanying adhesive sheets and conductive paste (Ten20 Conductive Paste, Weaver & Co.), the 8x8 1cm-spaced HDsEMG electrode grids (HD10MM0808, OT Bioelettronica, Fig. 1) were attached to the leg such that the centre of the arrays aligned visually with the muscle belly markings. A strip of hypoallergenic kinesiology tape was placed over the array to improve adhesion, and four reference fiducials were adhered directly to the skin surrounding the array. Once the fiducials had been attached, an elastic compression wrap was placed over the array to maximize the arrays' compliance to the skin and house the wireless amplifier. The array was then connected to the amplifier as well as the relevant ground electrode.

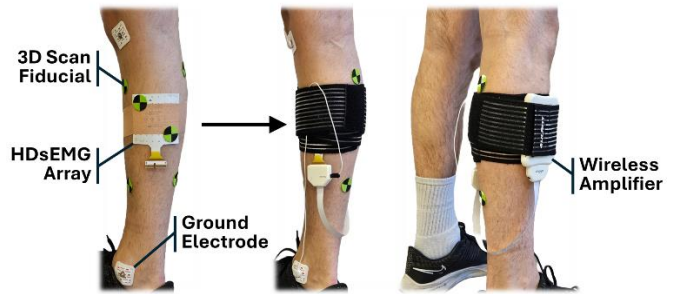

Fig. 1. Visual representation of the muscle instrumentation setup, specifically for the GM

### B. Experimental Protocol

Due to the size of the arrays, we could only instrument one muscle from the thigh and one from the shank at a time. We randomly selected muscles initially and repeated our protocol for the other muscles.

Following instrumentation, the participants were instructed to perform a selection of exercises, starting with 3x10-second repetitions of the isometric exercises specific to the currently instrumented muscles (Fig 2a-d), followed by 30 seconds worth of sit-to-stand (STS) repetitions and a 10-step timed-up-and-go (TUG) (Fig 2e-f), two dynamic exercises seen commonly during stroke assessment sessions [45], [46]. The HDsEMG exercise data was collected at a sample rate of 2kHz. Once the exercises had been completed, the elastic wrap was removed and a 3D scan of the leg with the fiducials still attached was taken using a 3D scanner (Structure Sensor Pro, Structure) attached to an iPad (iPad Pro 5$^{th}$-Generation, Apple).

After the scanning was complete, we removed the HDsEMG array and shifted it randomly away from the muscle belly and rotated it randomly about its center by -30 to 30 degrees before repeating the exercises and scanning. This was done three times, shifting the array to region 1-3 (Fig. 3a) to cover a range of possible shift scenarios. Once all data were collected, we removed the array and performed a final scan with just the leg and reference fiducials present.

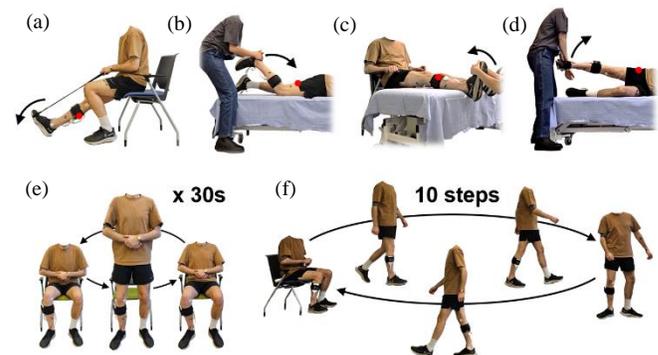

Fig. 2. Exercises performed by the participants, a) GM isometric, b) ST isometric, c) TA isometric, d) TFL isometric, e) 30-second sit-to-stand, f) 10-step timed-up-and-go

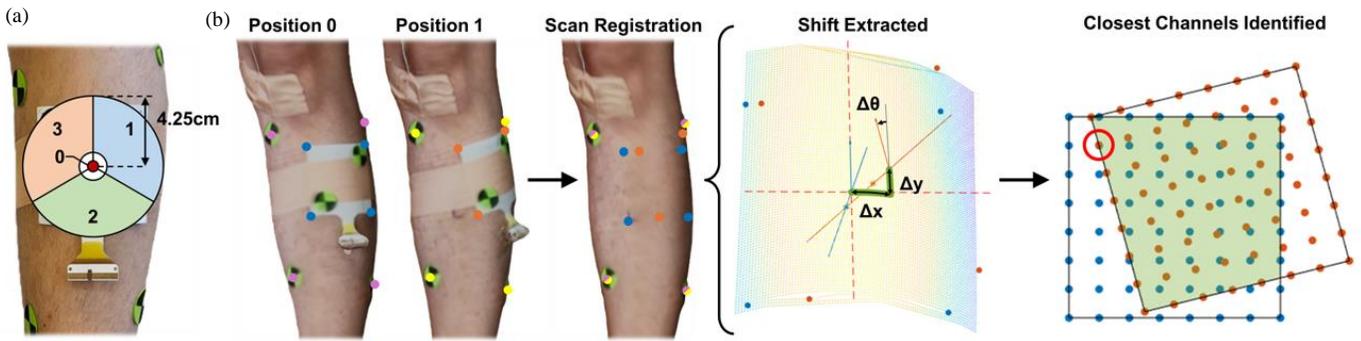

Fig. 3. a) Array shifting process, b) ground truth array shift extraction and closest channel identification process

### C. 3D Scan Ground Truth Shift Extraction

We determined a ground truth shift for the electrode arrays by tracking reference fiducials and the array corners in three scans: one with the array in its initial location, one with the array in its post-shift location, and a third of the leg without the array. Firstly, the locations of the centers of the reference fiducials were manually identified and recorded for each of the three scans, and the locations of the array's corners were identified and recorded for the two scans in which the array was present. Then, using the reference fiducial location information, both array scans were registered to the no-array scan. At this point, a subsection of the no-array scan was taken, retaining only the region bounded by the arrays' outer corners. Using the four corners of each array, the arrays' centers were identified and projected onto the surface of the scan. Finally, the transformation between the pre- and post-shift scans' array centers in terms of 'x', 'y', and 'θ' were extracted where 'x' is the distance over the surface of leg parallel to the top and bottom edges of the starting array, 'y' is the distance over the surface of the leg parallel to the vertical edges of the starting array, and 'θ' is the required rotation about the center of the array after the 'x-y' translation to map the array corners onto those post-shift (Fig. 3b). By understanding the physical shift that occurred, this facilitated the identification of the closest electrodes pre- and post-shift, as well as the surface distance between any two electrodes pre- and post-shift which will be used later in the data analysis.

### D. HDsEMG Feature Extraction

All HDsEMG data were first filtered between 20Hz and 450Hz using an $8^{th}$ order Butterworth bandpass filter before removing the 60Hz power line interference and its harmonics using a series of 2Hz-wide $2^{nd}$-order IIR notch filters. For each recording, any open channels were identified and removed from subsequent analysis; these were channels that didn't show any signal for either part of or the full recording. 82.3% had no open channels, 12.8% had a single open channel, while the most open channels in a single recording was 5.

Once filtered, the EMG signals were rectified and low-passed at 2Hz again using an $8^{th}$ order Butterworth filter to generate the signal envelopes before being segmented. For the isometric recordings, these were split into individual contractions by thresholding the signal envelope at >20% its maximum, while for the dynamic recordings the STS data were manually segmented into the individual stand-sit events, and the TUG data were manually segmented to retain only the peak activations within each individual step.

Once processed, we extracted six features from the HDsEMG segments. Firstly, we extracted the mean, median, and peak frequencies that, given they are known to differ between healthy and stroke populations, have been used in previous work to predict stroke [47]. Furthermore, a reduction in the mean and median frequencies specifically has being associated with increased muscle fatigue, something common in stroke patients [30]-[32], [48]. Given these features are extracted from the signals' power spectra, the decision was made to also inspect total spectral power. Then, integrated EMG (iEMG), reflecting voluntary muscle drive, was included given that increases in iEMG can suggest improvements in firing rate and motor unit recruitment, both of which can be linked to stroke recovery [49]-[51]. Finally, we also examined the signal envelope's maximum amplitude, as changes in this value have been shown to correlate with muscle force production, and thus may indicate improvements or degradation the patient's abilities [52], [53].

For the extraction of the frequency domain features, the power spectral density (PSD) as a function of frequency (f), and contraction number (c), S(f,c), was estimated using Welch's averaged periodogram with a 200ms Hamming window. The total power of each contraction, P(c), was then estimated through the numerical integration of the PSD as shown by equation 1.

$$P(c) = \sum_{f=0}^{Fs/2} S(f,c) \Delta f \quad (1)$$

The median frequency for each contraction, MDF(c), measured in Hz, was defined as the frequency at which 50% of the total power within the contraction's power spectrum was reached; this is shown by equation 2.

$$\sum_{f=0}^{MDF(c)} S(f,c)\Delta f = 0.5\, P(c) \quad (2)$$

The mean frequency for each contraction, MNF(c), was taken as the frequency in Hz where the average power across the spectrum was reached as calculated using equation 3.

$$MNF(c) = \frac{\sum_{f=0}^{Fs/2} f S(f,c)\Delta f}{\sum_{f=0}^{Fs/2} S(f,c)\Delta f} \quad (3)$$

As for the peak frequency, this was simply taken as the frequency value which showed the highest power within the power spectrum of each contraction (Fig. 4). Finally, for the time domain features, both the integrated EMG values and max envelope amplitude were extracted from the signals' envelopes: the integrated EMG value for each contraction was taken as the numeric integral of said contraction, while the max envelope amplitude was simply the greatest value of the envelope within a given contraction. Then, for all features, the final values used for comparison across recordings were computed as the contraction-averaged feature value (equation 4) where $N_c$ is the number of contractions identified within the recording. Note that for our time domain features, no signal normalization was performed.

$$\bar{F} = \frac{\sum_{c=1}^{N_c} F(c)}{N_c} \quad (4)$$

### D. Feature Analysis

Following the extraction of the six unique features, the subsequent analysis involved three main components: 1) exploring the difference in feature values with respect to distance intra-recording, 2) exploring the difference in feature values between the closest channels pre- and post-shift as identified using the methodology of Fig. 3, and 3) exploring the difference between the same channels pre- and post-shift, seeking to understand how these differences compare to those seen intra-recording.

For analysis 1), all possible electrode pairs within the grid were analysed (64^2, 4096 comparisons per recording), with the distance between them being the physical distance over the grid given its 8x8 configuration and 1cm inter-electrode spacing. For each channel pair across recordings, the absolute percent difference from the start electrode to the end electrode was computed to generate a relationship between distance and percent metric difference and a continuous function (inverse exponential) was fit to the data (example in Fig. 5).

Analysis 2) followed a similar procedure, comparing the closest electrodes within each shift pair. These were defined as being the results at the same location post-shift, although in actuality the mean distance between these closest electrodes as identified from the 3D scans was 0.12±0.10cm. We again computed the percent metric difference between same-location post-shift electrodes and used the median across all pairs to establish a baseline expected reapplication error (Fig. 5). It should be noted that given these were not the same electrodes pre- and post- shift, any differences seen likely arise as a combination of changes in the skin-electrode interface, one example being the variations in conductivity that will have arisen due to how the electrodes were prepared (e.g. some electrodes may have more conductive gel than others).

Finally, for analysis 3), in an attempt to separate the effects of reapplying the array and changing the recording electrode as seen in analysis 2), the channels used here for comparison were the two closest channels pre- and post- shift. For example, assuming that channel 19 pre-shift was closest to channel 23 post-shift, the present analysis would compute the feature difference between channel 19 pre- and post-shift, as well as channel 23 pre- and post-shift, with the distance between the electrodes extracted from the 3D scans. We again computed percent differences between each feature, before comparing these to the values from analysis 1) to understand how the inter-session trends differ from those seen intra-session (Fig. 5).

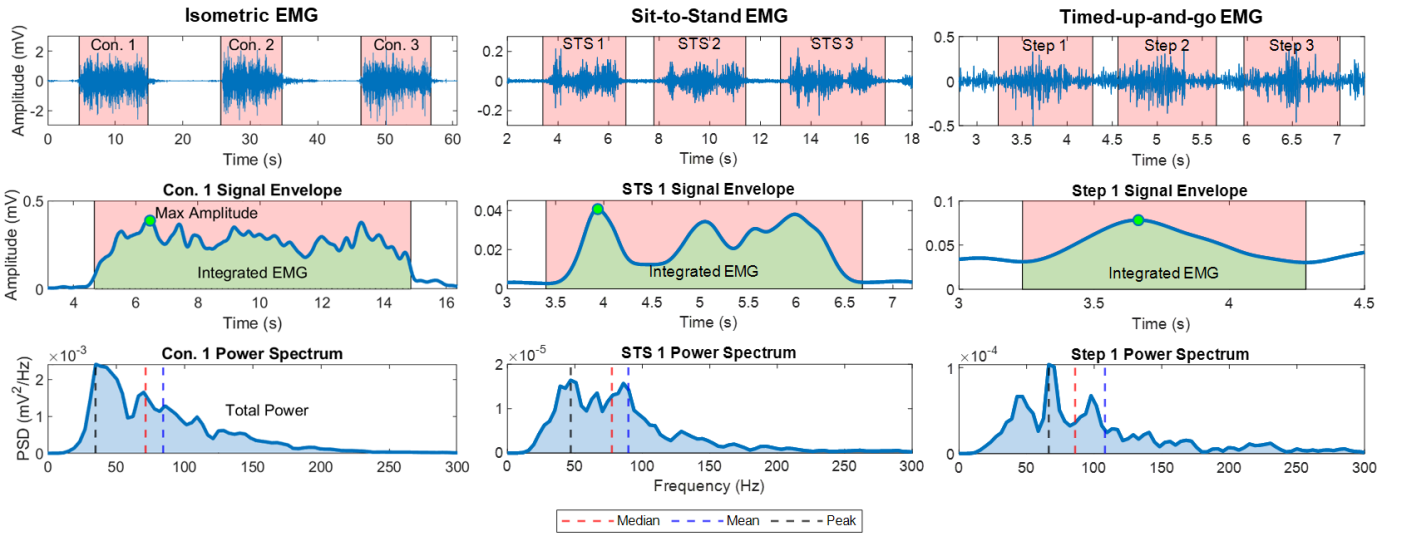

Fig. 4. Visual representation of the feature extraction process for a single segment of an isometric, sit-to-stand, and timed-up-and-go trial from the gastrocnemius medialis (GM)

## III. RESULTS

First, we present how the EMG features' values changed due to intra-recording electrode distance over the different muscles in the different activities, as well as the median same-location inter-recording differences (Fig. 6). Addressing analysis 1), the results indicate that the percent difference between EMG features increases with electrode distance intra-recording. Additionally, the curve shapes generally resembled an inverse exponential. The three frequency features saw the lowest absolute percent difference at a distance of 10cm for their combined muscles and exercises (14.4% for mean frequency, 16.0% for median frequency, and 14.9% for peak frequency). The largest 10cm absolute percent difference was seen within the total power (combined muscle and exercise difference of 56.8%). The TFL specifically had percent differences of 99.7% while also being the sole results that did not follow the inverse exponential trend.

While the magnitude of the differences vary across the features, we do still see some consistent trends. In the STS trials, the TA muscle presented the greatest percent difference across features, and the GM the lowest. In the TUG trials, the ST presented the greatest percent difference and the TFL the lowest, while for the ISO trials no single muscle consistently had the highest or lowest percent difference across the features.

Addressing analysis 2), we observed that the median inter-session, closest-electrode percent differences across features

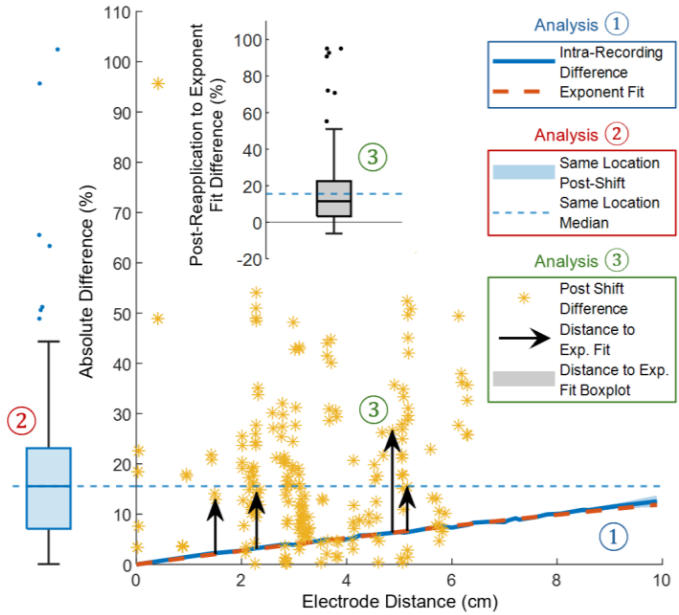

Fig. 5. Visualization of the feature analysis procedure for the integrated EMG values coming from the GM across STS trials where the shaded region surrounding the intra-recording difference of analysis 1 shows the 95% confidence interval.

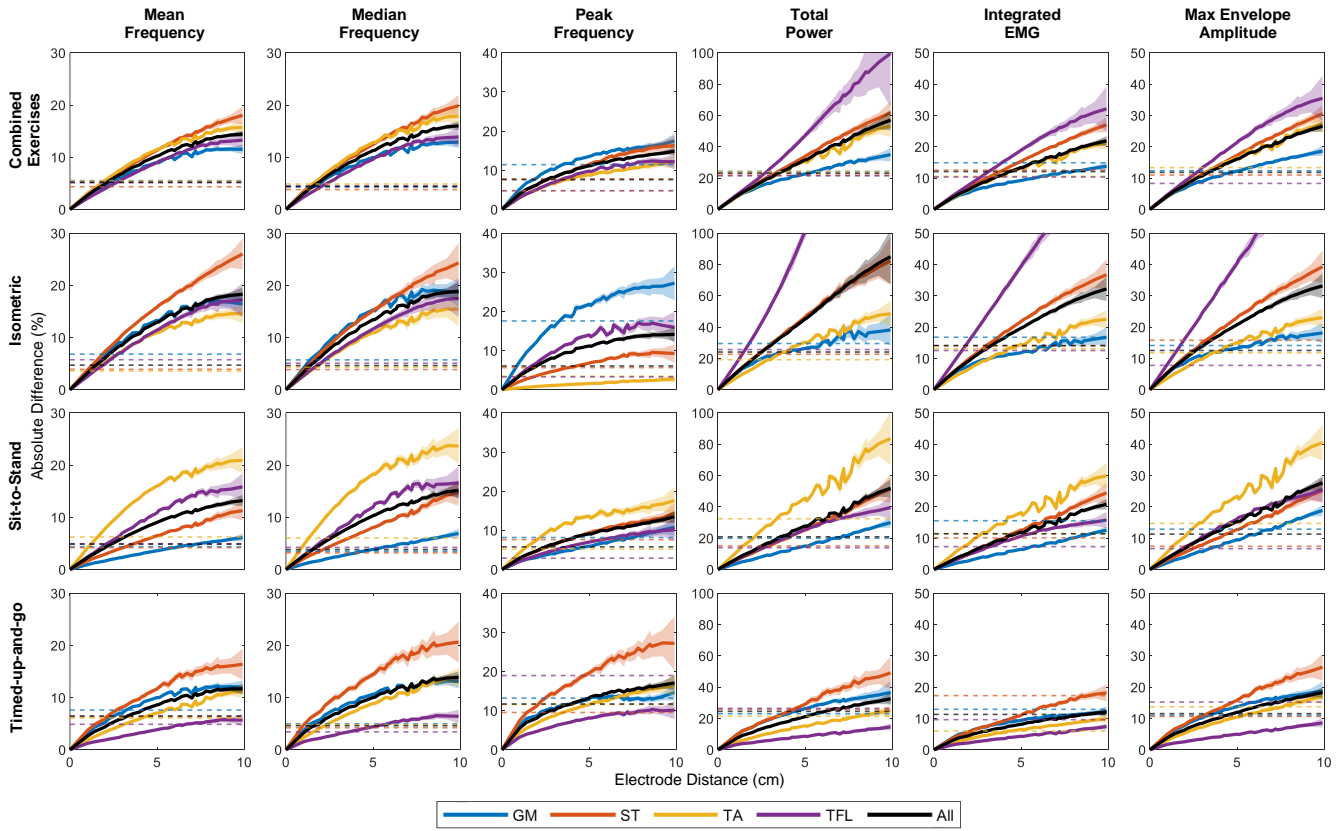

Fig. 6. Mean percentage absolute difference of each feature-exercise combination as a function of distance intra-recording. Shaded regions indicate 95% confidence interval while the dotted lines show the median same location post-shift difference.

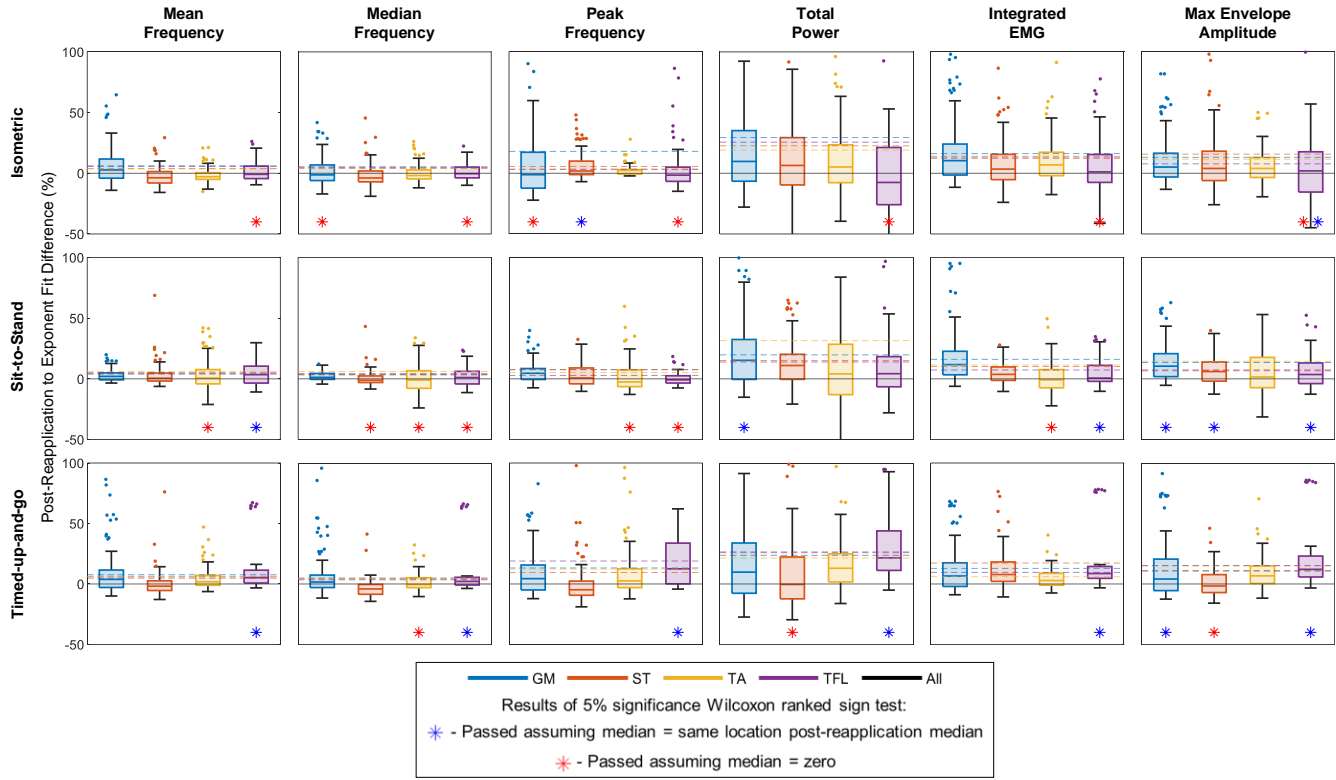

Fig. 7. Boxplots showing the difference in percent between the exponent fits and the same electrode inter-session feature differences. Asterisks represent the results of the Wilcoxon signed rank test at 5% significance: red asterisks show distributions that passed the test assuming zero median, blue asterisks are distributions that passed assuming a median equal to the median same location post-shift difference

were consistently <30% for all muscles and activities with the exception of the TA's total power in the STS trials (31.4%). Notably the mean and median frequency had the smallest median percent same-location differences for the combined exercises (<5.5%), while the total power had the greatest for the combined exercises (>20%).

Addressing 3), we explored how the percent differences due to the electrode distance (from analysis 1)) and the percent difference due to the inter-session recordings (from analysis 2)) compounded. To accomplish this, we subtracted the error from the inverse exponential intra-session fit from the percent error of the same-electrode inter-recording results at their given difference in location; this was then compared to the inter-session same-location percent errors. Here, we found that the median difference errors were consistently lower than the median inter-session errors. In some cases, seen mostly for mean, median, and peak frequency, the difference errors could be approximated as having a median of 0 (Fig. 7, red asterisk), whereas others, particularly the max envelope amplitude, could be approximated as having the same median as the inter-session same-location results (blue asterisk). In the case of these red asterisks, this implies that the feature differences with respect to distance approximate to the results seen intra-recording, while the blue asterisks imply that it's feature difference with respect to distance approximates to the intra-recording relationship shifted up by the same location median.

Across the results, there were some cases that had significant negative medians (TFL for total power, and ST frequently). However this could be due to curve fitting limitations (Fig. 6 for the TFL) Otherwise, the median differences all exceeded zero, but in no cases did the median exceed that of the same location difference, implying that this represents the maximum expected error when comparing features across recording sessions.

Finally, to obtain a broader understanding of how electrode distance intra-session relates to the same location median and better quantify individual muscles' robustness to variations in recording location, we calculated the percentage of features' differences less than the same location median at each distance (Fig. 8). Here, we found that across all exercises, electrode distances of 1cm result in 12-21% of feature differences exceeding the difference that would arise from sensor array reapplication (Fig. 8). The TFL for the timed-up-and-go was seen to be most robust where recording sites >10cm apart resulted in only 20% of feature differences exceeding inter-session differences. The TFL in the isometric and sit-to-stand exercises was more sensitive however, with 80% of feature differences being greater than the same-location inter-session differences at an electrode distance of 10cm.

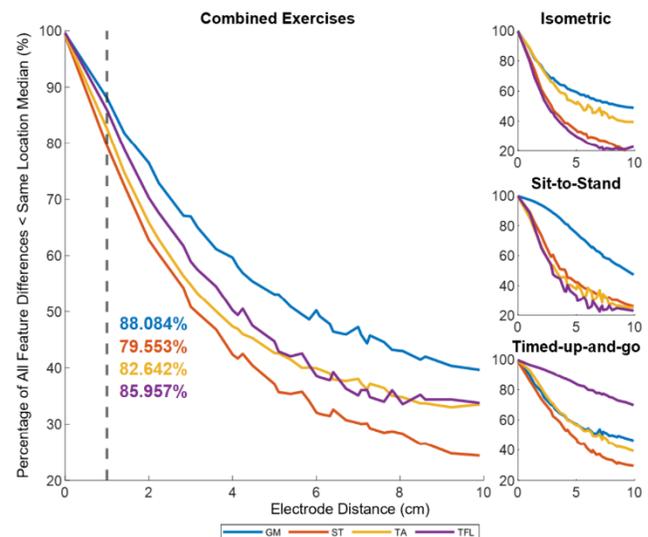

Fig. 8. Plot showing the percentage of all intra-recording feature differences less than the same location median for each exercise

## IV. Discussion

Within this study, we have presented a detailed analysis of a number of sources of error within sEMG measurement, concluding that the inter-session error is equal to the sum of the errors associated with the electrode distance and sensor reapplication (Fig. 7). We can say this given that, with the use of HDsEMG, we were able to isolate feature variabilities caused by the distance (controlling for reapplication errors) and variabilities caused by the reapplication (controlling for distance). By then comparing the intra-session distance and inter-session same location results of Fig. 6, we see that the dominant errors were those associated with distance, particularly when the distance is in excess of 3-6cm depending on the muscle of interest as reinforced by Fig. 8. Additionally, as supported by Fig. 7, we can say that should we be able to account for electrode shift inter-session, the maximum expected error will be equal to the median same-location error – or in other words the compound effect of the change in recording electrode and array reapplication.

Now, while this was seen to be true across features, those of the frequency domain were most robust to all sources of error investigated, likely down to the fact that they are not impacted by signal amplitude attenuation like the time domain features. Furthermore, per Fig. 7, these frequency domain features could most often be approximated as zero-median, and this coupled with their comparatively low variance implies that their feature difference trends intra- and inter-session were similar, making them better suited for direct comparison across recording sessions than the other features explored in this study. Conversely, the total power was seen to exhibit the greatest errors in each case, and was also the sole feature whose intra-session distance trend (albeit just for the TFL) did not follow an inverse exponential relationship, making it ill-suited to cross-session comparisons. Finally, while the iEMG and max envelope values saw greater errors across the board than the frequency features, unlike the total power there were some notable and fairly consistent trends. In particular, we saw how the max envelope value in Fig. 7 could often be expressed as a distribution centred around the same-location inter-session median, implying that the inter-session feature difference as a function of distance is equal to that seen intra-session shifted upwards by the same-location inter-session median.

This is significant given that, previously, this time domain feature required normalization by the maximum voluntary contraction (MVC) value in order to be compared across recording sessions. While this is acceptable for cases where the MVC remains stable across these sessions, for a stroke population where changes in muscle capacity are not uncommon, it would be beneficial to be able to compare the envelope amplitudes more directly. To this point, the results of Fig. 7 show that this is possible so long as we can account for where we are recording from with respect to the initial recording site, and that we know what the median reapplication error is. Up to this point however, all of the absolute feature differences have been expressed as a percentage to facilitate cross-feature comparisons, although in reality it is oftentimes more intuitive to compare envelope amplitudes directly to improve our understanding of motor unit recruitment and force production, while also allowing for cross-patient comparisons. In response, the max envelope amplitude results in volts are presented below within Fig. 9.

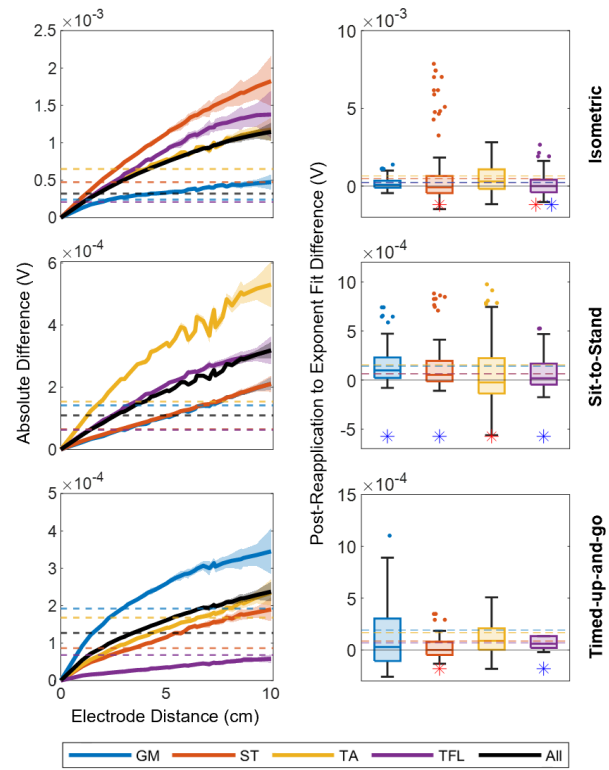

Fig. 9. Max envelope amplitude results presented in volts

From this figure, we see that the inverse exponential trend with respect to distance intra-session remains and that the maximum median same-location thresholds for the isometric and dynamic exercises were 0.65mV and 0.19mV respectively. Thus, so long as we record from the same location across recording bouts, should the median of the inter-session differences exceed these values it can most likely be explained by physiological changes in the test subject.

Beyond the individual features, it is important to acknowledge the trends seen with respect to the muscles and exercises included in our analysis. On the point of the exercises, we saw from Fig. 6 that generally the distance errors were lower in the dynamic exercises than the isometrics, with one theory as to why this may have been being the fact that the dynamic exercises result in the movement of the muscle underneath the sensing array. This could have reduced the effect of distance given that the same electrode would have seen the response from a variety of locations on the muscle, thus having an almost averaging-like effect on the features. In terms of the muscles, we saw that, through Fig. 8, consistently the GM was most robust to distance errors with the exception of the TFL for the TUG data. Although we cannot say why exactly this was the case, we theorize that due to the TFL's low activation during walking, particularly for healthy subjects, the signal is comprised primarily of activity from the surrounding muscles propagating to the recording site, and thus the effects of placement variations are reduced.

To end, it is important to acknowledge the limitations of this study. Firstly, given that the array reapplication was performed across the span of a single session, the array was not reprepared with fresh conductive paste between recordings. As such, we cannot say for certain whether this would have a bearing on the results; on one hand it is possible that fresh paste would mitigate the errors seen when comparing features extracted from different electrodes, but on the other hand recording on separate days would likely incur additional errors due to changes in the skin electrode interface due to variations in additional parameters such as hair length and sweat level if the heat or humidity has changed. Additionally, analysis 2) centered around exploring the errors when recording from the same location inter-session, although given the approach we took the electrodes were not quite in the exact same place, although these distances were still well-within the 1cm inter-electrode distance. Finally, it is important to note that despite the motivation behind this work being stroke assessments, the results presented here came from a healthy population and thus we cannot say whether these trends would generalize to a stroke population.

## V. Conclusion

Through this study, we have been able to isolate the inter-session effects of electrode shift and sensor reapplication, the latter relating to changes in the properties of the skin electrode interface. In doing so, we were able to first conclude that, of these, electrode shift is the dominant error, although there are still errors seen when recording from the same location post-shift given the array reapplication process and changes to the properties of the recording electrode (e.g. its conductivity). Across the features explored in the study, this median same-location error represented the maximum expected error when controlling for distance, and thus seeing a change in feature value exceeding this can indicate physiological changes in the individual. Furthermore, we saw the lowest reapplication errors in the frequency features, meaning that their trends with respect to shift distance inter-session can be approximated as that intra-session, while the trends for the time domain features (particularly max envelope amplitude) can instead be approximated as the intra-session trend shifted by the median same-location error. Regardless, these results highlight that should we be able to control for electrode shift inter-session, we can more directly compare sEMG features indicative of stroke recovery over time, furthering the technology's utility in the field of remote clinical assessment.

## Acknowledgment

The authors would like to acknowledge the contributions of Ryan Lo for his assistance in data segmentation. The authors would also like to acknowledge funding from the Canada Foundation for Innovation, the Canadian Natural Sciences and Engineering Research Council, and AGE-WELL. They have no conflicts of interest to disclose.

## References

[1] M. Kazamel and P. P. Warren, "History of electromyography and nerve conduction studies: A tribute to the founding fathers," *Journal of Clinical Neuroscience*, vol. 43, pp. 54–60, Sep. 2017, doi: 10.1016/J.JOCN.2017.05.018.

[2] M. B. I. Reaz, M. S. Hussain, and F. Mohd-Yasin, "Techniques of EMG signal analysis: detection, processing, classification and applications," *Biol Proced Online*, vol. 8, no. 1, p. 11, Mar. 2006, doi: 10.1251/BPO115.

[3] E. R. Avila, S. E. Williams, and C. Disselhorst-Klug, "Advances in EMG measurement techniques, analysis procedures, and the impact of muscle mechanics on future requirements for the methodology," *J Biomech*, vol. 156, p. 111687, Jul. 2023, doi: 10.1016/J.JBIOMECH.2023.111687.

[4] M. Al-Ayyad, H. A. Owida, R. De Fazio, B. Al-Naami, and P. Visconti, "Electromyography Monitoring Systems in Rehabilitation: A Review of Clinical Applications, Wearable Devices and Signal Acquisition Methodologies," *Electronics 2023, Vol. 12, Page 1520*, vol. 12, no. 7, p. 1520, Mar. 2023, doi: 10.3390/ELECTRONICS12071520.

[5] S. H. Roy *et al.*, "A combined sEMG and accelerometer system for monitoring functional activity in stroke," *IEEE Transactions on Neural Systems and Rehabilitation Engineering*, vol. 17, no. 6, pp. 585–594, Dec. 2009, doi: 10.1109/TNSRE.2009.2036615.

[6] K. Kisiel-Sajewicz *et al.*, "Weakening of synergist muscle coupling during reaching movement in stroke patients," *Neurorehabil Neural Repair*, vol. 25, no. 4, pp. 359–368, May 2011, doi: 10.1177/1545968310388665/ASSET/IMAGES/LARGE/10.1177_1545968310388665-FIG5.JPEG.

[7] J. J. Eng and A. M. Pastva, "Advances in Remote Monitoring for Stroke Recovery," *Stroke*, vol. 53, no. 8, 2022, doi: 10.1161/STROKEAHA.122.038885.

[8] E. Appleby, S. T. Gill, L. K. Hayes, T. L. Walker, M. Walsh, and S. Kumar, "Effectiveness of telerehabilitation in the management of adults with stroke: A systematic review," *PLoS One*, vol. 14, no. 11, p. e0225150, Nov. 2019, doi: 10.1371/JOURNAL.PONE.0225150.

[9] P. Webster, "Virtual health care in the era of COVID-19," *The Lancet*, vol. 395, no. 10231, pp. 1180–1181, Apr. 2020, doi: 10.1016/S0140-6736(20)30818-7.

[10] J. Wosik *et al.*, "Telehealth transformation: COVID-19 and the rise of virtual care," *Journal of the American Medical Informatics Association*, vol. 27, no. 6, pp. 957–962, Jun. 2020, doi: 10.1093/JAMIA/OCAA067.

[11] G. Ebenbichler, K. Kerschan-Schindl, T. Brockow, and K. L. Resch, "The future of physical & rehabilitation medicine as a medical specialty in the era of evidence-based medicine," 2008. doi: 10.1097/PHM.0b013e31815e6a49.

[12] E. Haeuber, M. Shaughnessy, L. W. Forrester, K. L. Coleman, and R. F. Macko, "Accelerometer monitoring of home- and community-based ambulatory activity after stroke," *Arch Phys Med Rehabil*, vol. 85, no. 12, pp. 1997–2001, Dec. 2004, doi: 10.1016/J.APMR.2003.11.035.

[13] D. Rand, J. J. Eng, P. F. Tang, J. S. Jeng, and C. Hung, "How Active Are People With Stroke?," *Stroke*, vol. 40, no. 1, pp. 163–168, Jan. 2009, doi: 10.1161/STROKEAHA.108.523621.

[14] D. Rand and J. J. Eng, "Disparity between functional recovery and daily use of the upper and lower extremities during subacute stroke rehabilitation," *Neurorehabil Neural Repair*, vol. 26, no. 1, pp. 76–84, Jan. 2012, doi: 10.1177/1545968311408918.

[15] K. S. Hayward, J. J. Eng, L. A. Boyd, B. Lakhani, J. Bernhardt, and C. E. Lang, "Exploring the Role of Accelerometers in the Measurement of Real World Upper-Limb Use After Stroke," *Brain Impairment*, vol. 17, no. 1, pp. 16–33, Mar. 2016, doi: 10.1017/BRIMP.2015.21.

[16] S. Halloran, L. Tang, Y. Guan, J. Q. Shi, and J. Eyre, "Remote monitoring of stroke patients' rehabilitation using wearable accelerometers," *Proceedings - International Symposium on Wearable Computers, ISWC*, pp. 72–77, Sep. 2019, doi: 10.1145/3341163.3347731.